# A Computational Study Explaining Processes Underlying Phase Transition

**Sasanka Sekhar Chanda**

*Indian Institute of Management Indore*

Email: sasanka2012@gmail.com

**Bill McKelvey**

*University of California Los Angeles*

Email: bill.mckelvey@anderson.ucla.edu

---

**Abstract.** In real-world systems, phase transitions often materialize abruptly, making it difficult to design appropriate controls that help uncover underlying processes. Some agent-based computational models display transformations similar to phase transitions. For such cases, it is possible to elicit detailed underlying processes that can be subsequently tested for applicability in real-world systems. In a genetic algorithm, we investigate how a modest difference in the concentration of correct and incorrect knowledge leads to radically different outcomes obtained through learning efforts by a group of agents. We show that a difference in concentration of correct and incorrect knowledge triggers virtuous and vicious cycles that impact the emergent outcome. When virtuous cycles are in operation, <u>delaying</u> the onset of equilibrium attains superior outcomes. For the vicious cycles, <u>reaching</u> equilibrium <u>quickly</u> attains superior outcomes. Our approach helps uncover simple mechanisms by which Nature works, jettisoning the yoke of unrealistic assumptions endemic in mathematics-based approaches. Our explanatory model helps direct research to investigate concentrations of inputs that obtains outcomes on the favourable side of phase transitions. For example, by tracking change in concentration of relevant parameters, scientists may look for reasons why cells cease to reproduce fit cells in organs. This can help design rejuvenation of organs. Further, in the world of physics, our model may inform in situations where the dominant Ising model falls short.

---

# Introduction

In phase transition phenomena, a system displays a persistence of significantly different values for an outcome of interest, when certain input conditions differ. For our purposes, a system is an embodiment of the interplay of micro-level processes from which a macro-level outcome of interest emerges. Even outside the heavily-researched areas of chemistry, particle physics, cosmology and earth sciences, phase transition phenomena are ubiquitous in society[1], economics[2], technology[3, 4] and biology[5].

- ❖ An underdeveloped or ravaged country – South Korea or post-war Japan in the fifties – develops into an industrial behemoth.
- ❖ Thriving economies of South East Asia, like Indonesia and Malaysia became a shadow of their former selves after the ASEAN crisis in the late nineties.
- ❖ Telecommunication towers that send and receive cell phone signals radically underperform upon the materialization of certain changes in geographical distribution of mobile subscribers during the day.
- ❖ A cell loses its ability to divide at a certain point; an organ ceases to be vitally active if a critical mass of cells lose their ability to divide. And so on.

In physical phenomena, phase transitions tend to materialize rapidly[6]. This makes it difficult to design appropriate controls to study phase transition mechanisms. Computational simulation experiments provide an opportunity to carry out in-depth study of mechanisms by which phase transition occurs[7]. In our research we highlight a situation akin to phase-transition, and delve into the biological process by which it occurs. Using a genetic algorithm, we explain how a modest difference in the proportions of correct and incorrect knowledge in agents leads to radically higher (or lower) group-level knowledge outcomes obtained through learning efforts by the agents.

Our research provides a hypothetical phase-transition process that can be tested for its applicability in a variety of social, economic, and biological phenomena. Where tests are positive, systems may be designed to operate on the favourable side of a phase transition. In general, phase transition processes elicited from computational experiments may be tested for applicability in several interesting natural phenomena, departing from extant approaches using mathematical functions. Thereby, scientific inquiry is freed from constraints arising from having to incorporate unpalatable assumptions for the sole sake of mathematical tractability, i.e., biological phenomena don't have to be treated like the physical phenomena that physicists have studied for centuries and about which mathematics has been historically created. This creates a more realistic pathway for the discovery of the various ways in which Nature works.

**Simulation model**

Suppose we, and readers, are interested in finding the values in an *M*-bit string that we call "**R**" (e.g., **R** for *Reality*), where each bit (or dimension) takes values from the set $S1 = \{-1, +1\}$ with probability 0.50. Suppose further, that we use computational simulations of a genetic algorithm, to find values in as many bits of **R** as possible. Accordingly, we generate a population of *N* agents. Each agent is an *M*-bit string, where each bit takes values from the set $S2 = \{-1, 0, +1\}$ with equal (one-third each) probability. We call this population a *Marchian* population, in honour of James March, who provided the specifications for our genetic algorithm[8]. Also suppose that we have one more *M*-bit string, named *Group Code* – abbreviated as **GC** – where we compile the guesses regarding values in **R**. Initially, all bits of **GC** are given value "0". The knowledge of **GC** (or any given agent) is computed as the number of matches between values in bits of **R** with values in corresponding bits of **GC** (or the agent in consideration) divided by *M*. In each time-step of a simulation, **GC** learns from the agents and the agents learn from **GC**.

In the computational model, **GC** is given an ability to identify which agents (who are called "elites" below) know more about the reality **R** than itself. In each period, **GC** forms a team of such elites from the agents in the group. For each bit, a vote is taken among the elites, assessing how many are in favour of "+1" and how many are in favour of "–1". Suppose the difference is *K* units. When $K > 0$, **GC** updates its value to that preferred by the elites with probability $[1 – (1– p_2)^K]$ where $p_2$ is the learning rate of **GC** ($0 < p_2 < 1$).

In order to learn from **GC**, in each period, for each bit position, an agent looks at the value in the corresponding bit-position of **GC**. If the value in **GC** is non-zero and different from that in the corresponding position in the agent's string, the agent updates its value in that dimension with a probability $p_1$ ($0 < p_1 < 1$). Equilibrium is reached when values in **GC** and values in all agents converge.

## Results

In the first set of experiments all agents learn at the same rate, $p_1$. We vary $p_1$ and observe the level of knowledge in **GC** when a point is reached whereby **GC** and the agents have identical values in their knowledge dimensions and learning ceases (i.e. equilibrium is reached). The results are shown in line L1 of Table 1. The values reported are based on averages resulting from over 10,000 computational runs. The knowledge in **GC** is seen to decrease monotonically as $p_1$ increases. Thus, faster learning by agents erodes the heterogeneity of member knowledge; faster learning by agents causes them to come to agreement more quickly, which results in a lower level of shared knowledge.

**Table 1 | Equilibrium knowledge in Group Code**

|   | Agent population and learning rate heterogeneity | Average learning rate of agents ($p_{1\text{-AVG}}$) | | | | | | |
|---|---|---|---|---|---|---|---|---|
|   |   | 0.2 | 0.3 | 0.4 | 0.5 | 0.6 | 0.7 | 0.8 |
| L1 | *Marchian* population, all agents learn at the same rate | 0.800 | 0.771 | 0.748 | 0.731 | 0.717 | 0.707 | 0.698 |
| L2 | *Marchian* population, mix of slow and fast learners | 0.817 | 0.809 | 0.797 | 0.786 | 0.772 | 0.756 | 0.727 |
| L3 | *Sub-Marchian* population, mix of slow and fast leaners | 0.173 | 0.182 | 0.200 | 0.219 | 0.250 | 0.294 | 0.349 |

Parameters. $M = 30$, $N = 50$, $p_2 = 0.50$, 10,000 iterations

In a second set of experiments, two kinds of agents are deployed. *Slow Learners* learn at $p_1 = 0.10$; *Fast Learners* learn at $p_1 = 0.90$. In the results presented in line L2 of Table 1 we see that, for equivalent average rates of learning ($p_{1-\text{AVG}}$), a mix of slow and fast learners attains outcomes superior to that obtained by all agents learning at the same rate (line L1). In March's terminology, *exploration* is high to the left, since a larger proportion of slow learners allow knowledge heterogeneity (variance) to persist longer; *exploitation* is high to the right, since knowledge heterogeneity (variance) is quickly eroded when a larger proportion of agents learn fast.

In the third set of experiments, we create a *sub-Marchian* population by overwriting 15% of the randomly-selected bits of all agents with values opposite that of the value in the corresponding bit-position of **R**. Basically we increase the concentration of incorrect knowledge (held by the agents) relative to correct knowledge. All other parameters are exactly the same as in our second experiment. We observe a phenomenon akin to a phase transition when we compare outcomes from our second and third set of experiments. With *sub-Marchian* populations, the outcome (Line L3 in Table 1) is capped at 35%. In contrast, for *Marchian* populations (Line L2 in Table 1) the outcomes range from 73% to 82%, i.e., are 2+ times higher. We also note that the strategy for obtaining superior outcomes has changed: in line L2, higher exploration is preferable over higher exploitation; but in line L3, higher exploitation is preferable.

## Explanatory processes underlying phase transition

In order to find: (a) why outcomes shift drastically and (b) why strategies to obtain better outcomes – higher exploration for *Marchian* populations and higher exploitation for *sub-Marchian* populations – are different, we first look at the extent of correct and incorrect knowledge getting into **GC**, early in the simulation, as shown in Table 2.

**Table 2 | Correct and incorrect knowledge in the Group Code in early stages of simulation**

|    | *Population type* | *Time-Step→* | $t = 1$ | $t = 2$ | $t = 3$ |
|----|---|---|---|---|---|
| L1 | *Marchian population* | Correct knowledge in **GC** | 0 | 0.401 | 0.602 |
| L2 |  | Incorrect knowledge in **GC** | 0 | 0.402 | 0.317 |
| L3 | *Sub-Marchian population* | Correct knowledge in **GC** | 0 | 0.069 | 0.086 |
| L4 |  | Incorrect knowledge in **GC** | 0 | 0.839 | 0.867 |

Parameters. $M = 30$, $N = 50$, $p_2 = 0.50$, 10,000 iterations

We note that, by design, in any given time-step, **GC** learns from the previous time-step's representation of agent-knowledge values, and agents learn from the previous time-step's representation of **GC** knowledge-values. Thus, the values we see in the first three time-steps are those where **GC** learns from agents, where agents' knowledge have not changed from initial values. We also note that, a *Marchian* population initiates with about 33% correct knowledge and 33% incorrect knowledge. Thus, it is understandable that, in the beginning, a *Marchian* population pushes in roughly equal proportions (40%) of correct and incorrect knowledge into **GC** (as seen under $t = 2$; lines L1 and L2 of Table 2).

In contrast, *sub-Marchian* populations initiate with 28% correct and 43% incorrect knowledge. Due to this difference, in time-step two (column $t = 2$ in Table 2), **GC** gets about 84% incorrect knowledge (L4) and about 7% correct knowledge (L3). In the beginning, potentially all agents qualify as elites that advise **GC**, since **GC** has zero knowledge. Higher concentration of incorrect knowledge leads to a preponderance of incorrect knowledge in **GC**.

Furthermore, for time-step three, *Marchian* populations present a high threshold of 40% (line L1, column $t = 2$) for getting into the team of elites the **GC** learns from. The corresponding figure is about 7% for *sub-Marchian* population (line L3, column $t = 2$). Hence, for *sub-Marchian* populations, a far higher number of agents with faulty knowledge advise **GC**. This creates problems with respect to the eviction of faulty knowledge from **GC**, when it learns from *sub-Marchian* populations. In time-step three, we observe that whereas incorrect knowledge reduces from 40% to 31% in *Marchian* populations (L2), incorrect knowledge in **GC** actually increases to about 87% in *sub-Marchian* populations (L4). The contours of a phase transition are discernible as well. By time-step three, *Marchian* populations are seen to confer about 60% correct knowledge to **GC** (line L1, column $t = 3$). The corresponding figure is about 9% in *sub-Marchian* populations (line L3, column $t = 3$).

**Figure 1 | Change in agent knowledge over time, under high exploration, *Marchian* populations**

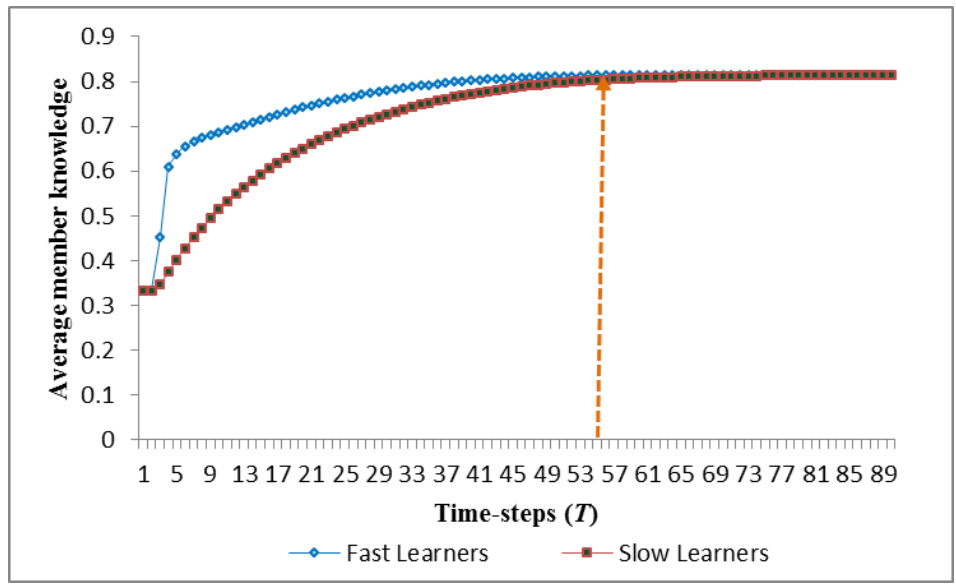

*Notes.* Rhombus-shaped markers represent Fast Learners who learn at a high rate ($p_1 = 0.9$). Rectangular markers represent Slow Learners who learn at a low rate ($p_1 = 0.1$). Other parameters are $M = 30$, $N = 50$, $p_{1-\text{AVG}} = 0.20$, $p_2 = 0.5$, Iterations = 10,000, *Marchian* population. The dotted arrow shows the point at which the two curves differ by less than one percent in value.

**Figure 2 | Change in agent knowledge over time, under high exploitation, *Marchian* populations**

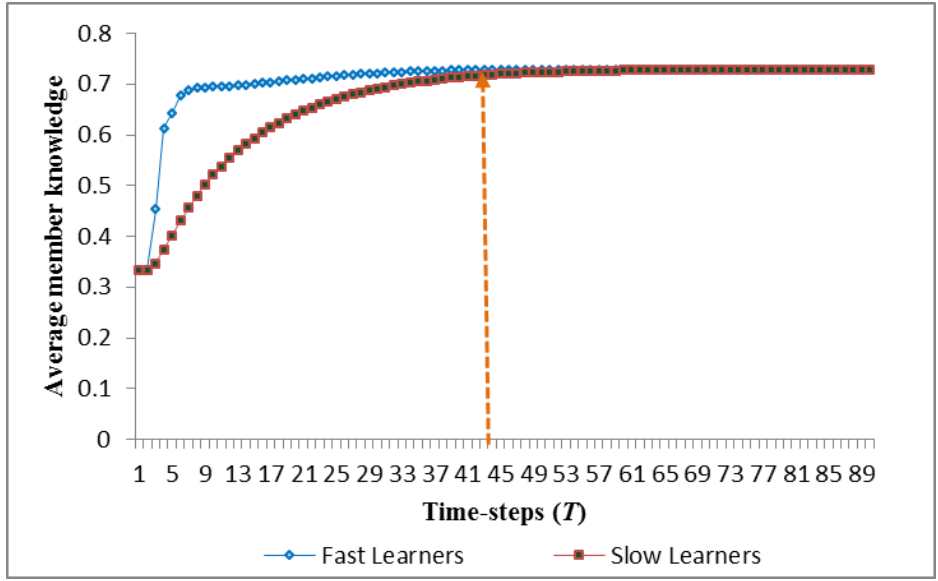

*Notes.* Rhombus-shaped markers represent Fast Learners who learn at a high rate ($p_1 = 0.9$). Rectangular markers represent Slow Learners who learn at a low rate ($p_1 = 0.1$). Other parameters are $M = 30$, $N = 50$, $p_{1-AVG} = 0.80$, $p_2 = 0.5$, Iterations = 10,000, *Marchian* population. The dotted arrow shows the point at which the two curves differ by less than one percent in value.

We now turn to explaining why the successful strategy of high exploration must change to that of high exploitation, i.e., from *Marchian* to *sub-Marchian* populations. In Figure 1, we present changes in agent knowledge over time, for high exploration ($p_{1-AVG} = 0.20$), for Marchian populations. Figure 2 presents corresponding information for high exploitation ($p_{1-AVG} = 0.80$). We observe that, in both cases, equilibrium sets in shortly after the knowledge of *Slow Learners* and *Fast Learners* attain same value. For high exploration (Figure 1), the convergence happens around time-step 55; for high exploitation the convergence occurs earlier, around time-step 44 (Figure 2). The catching-up happens earlier for high exploitation because the proportion of fast learners is higher. Thus, in *Marchian* populations, high exploration attains superior outcomes owing to knowledge heterogeneity of slow learners lasting for longer.

In Figure 3, we present changes in agent knowledge over time, for high exploration ($p_{1-AVG} = 0.20$), in *sub-Marchian* populations. Figure 4 presents corresponding information for high exploitation ($p_{1-AVG} = 0.80$). We observe that, in both cases, the knowledge of the *Fast Learners* decreases initially, given that they learn rapidly from the highly faulty **GC**. Note, though, by design, the knowledge of the *Fast Learners* cannot become less than that of **GC**. Hence, after a point, there is a turnaround. Knowledge of *Fast Learners* increases thereafter.

**Figure 3 | Change in agent knowledge over time, under high exploration, *sub-Marchian* populations**

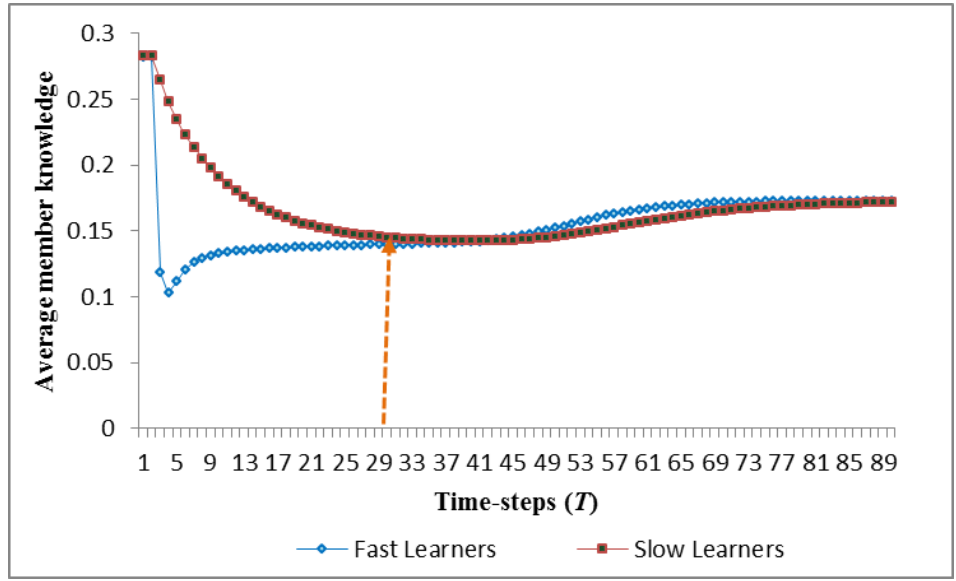

*Notes.* Rhombus-shaped markers represent Fast Learners who learn at a high rate ($p_1 = 0.9$). Rectangular markers represent Slow Learners who learn at a low rate ($p_1 = 0.1$). Other parameters are $M = 30$, $N = 50$, $p_{1–AVG} = 0.20$, $p_2 = 0.5$, Iterations = 10,000, *sub-Marchian* population with deficiency 15% from *Marchian* population. The dotted arrow shows the point at which the two curves differ by less than one percent in value at the time the fast learners catch up with slow learners.

**Figure 4 | Change in agent knowledge over time, under high exploitation, *sub-Marchian* populations**

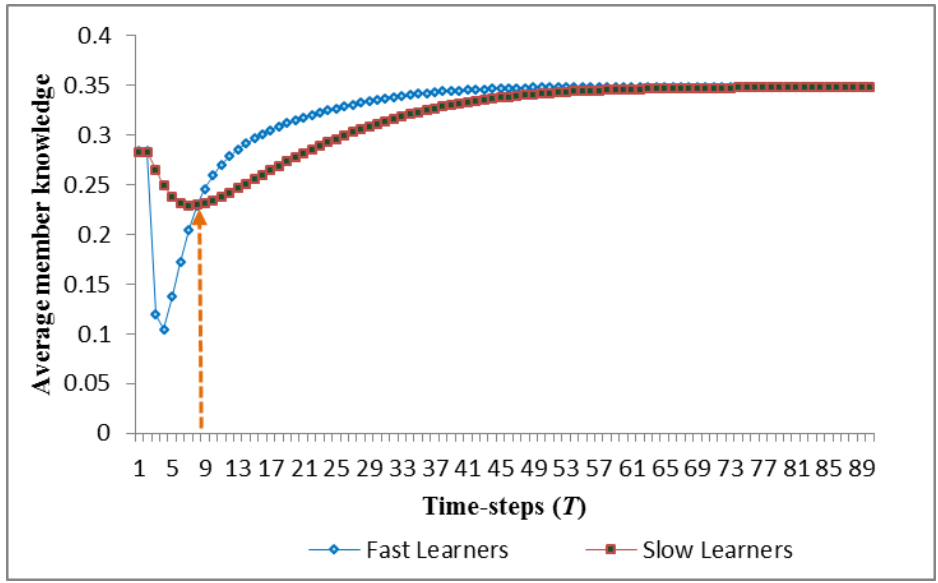

*Notes.* Rhombus-shaped markers represent Fast Learners who learn at a high rate ($p_1 = 0.9$). Rectangular markers represent Slow Learners who learn at a low rate ($p_1 = 0.1$). Other parameters are $M = 30$, $N = 50$, $p_{1–AVG} = 0.80$, $p_2 = 0.5$, Iterations = 10,000, sub-Marchian population with deficiency 15% from Marchian population. The dotted arrow shows the point at which the two curves differ by less than one percent in value at the time the fast learners catch up with slow learners.

Knowledge of the slow learners decreases as well, but at a lower rate, as expected. The *Fast Learners* catch up with the knowledge level of the *Slow Learners* by time-step 8 for high exploitation, and not before time-step 30 for high exploration. As in the previous case (pertaining to *Marchian* populations), the greater proportion of fast learners in high exploitation explains the quicker catch-up. The effect of quicker catch-ups on the eventual knowledge level of the **GC**, however, is different. For high exploration (Figure 3) the interception happens when the knowledge of the *Slow Learners* has degenerated to about 14%; for high exploitation, the corresponding value is 23% (Figure 4). This difference turns out to be crucial: High exploitation results in an eventual value of about 35% in **GC** (line L3, Table 1); high exploration confers about 18%.

All in all, for *Marchian* populations, the benefits of exploration over exploitation materialize through a higher proportion of *Slow Learners*, by having knowledge heterogeneity lasting for longer. In contrast, for *sub-Marchian* populations, the benefits of exploitation over exploration transpire by having higher proportion of *Fast Learners*, and by halting the loss of knowledge heterogeneity of *Slow Learners* earlier.

**Process summary**

In the case of a *sub-Marchian* population, the fact that the proportion of incorrect knowledge bits is higher than the proportion of correct knowledge bits results in the group code (**GC**) being populated with a high amount of incorrect knowledge, and a low amount of correct knowledge, right at the beginning. A high proportion of incorrect knowledge makes learning from **GC** toxic, and results in a loss of the knowledge heterogeneity of agents learning from **GC**. A low proportion of correct knowledge in **GC** exacerbates the situation, because the toxicity of **GC** is nurtured by having a low bar to **GC** being advised by agents with highly faulty knowledge. In this situation, sooner the vicious cycle is broken, the better. Thus, shorter exposure to the toxicity of **GC** is beneficial – as happens for high rate of exploitation – such that at least some part of the heterogeneity of agents is saved from destruction, and eventually contributes to knowledge of **GC**.

For a *Marchian* population, however, presence of high proportion of correct bits in **GC** early on leads to a high bar with regard to choice of agents that advise **GC**. Therefore, **GC** gets high quality advice, and this leads to flushing out of incorrect knowledge from **GC**. In this situation, the agents learning from **GC** also receive good quality knowledge. The longer the virtuous cycle continues, higher is the outcome obtained. Exploration occurs for longer and attains better outcomes compared to exploitation.

In a wide range of real-world systems, we observe that a modest difference in initial conditions is responsible for significant difference in equilibrium outcomes. Our exposition illustrates that the difference in initial concentrations of correct and incorrect knowledge leads to materialization of vicious and virtuous cycles (alternately termed as downward and upward spirals, respectively). A faster journey to equilibrium is beneficial when vicious cycles are in operation; a slower journey is beneficial when virtuous cycles are in operation. Gains from the upward spiral cumulated with the losses from the downward spiral account for the difference in outcome values on the two sides of the phase transition.

## Conclusion

Computational phase transition studies enable the delving into the micro-processes underlying a phenomenon of interest. This approach elicits detailed descriptive processes with far more ease than what is possible in studies of physical phenomena by physicists and others. These process descriptions allow the design of experiments to test whether a hypothesized process is applicable in a range of real-world non-physic types of systems. It becomes possible to design systems that operate on the favourable side of the phase transition. We hope that our work offers a method that helps scientific inquiry, just as the 'small-world' model of networks[9, 10] found applications in domains ranging from subcellular networks to ecosystems and the Internet[11, 12, 13].

The approach we suggest frees up the design of processes from being subverted to the incorrect assumptions endemic in mathematical modelling. The rate-equations of mathematics fail to handle irreversibility because left-hand and right-hand derivatives are not equal. Consequently, mathematical functions frequently fail to model complex, path-dependent interactions appropriately. In previous research, many of the results cited as 'anomaly', 'singularity' etc., are actually products of little-understood interactions and transformations of unrealistic assumptions arising from the tyranny of mathematical tractability. Prominent computational models like genetic algorithms[14], ant-colony optimization[15, 16], swarm intelligence[17], Kaufman's NK model[18], etc. provide excellent starting points to search for phase transition phenomena and inquire into underlying mechanisms. By using various agent-based model-designs of phase transitions, we are able to elicit simple processes to account for differences in macro-level outcomes by tracking changes in the distribution of micro-entities of interest. Thereby, we learn from computers, even so, we obtain insights into the inner workings of the more biological, economic, and social aspects of Nature.